**Dirac Fermion kinetics in three-dimensionally curved graphene**

By *Yoichi Tanabe[1*], Yoshikazu Ito[2*], Katsuaki Sugawara[3,4], Mikito Koshino[5], Shojiro Kimura[6], Tomoya Naito[7,8], Isaac Johnson[9], Takashi Takahashi[4], Mingwei Chen[4,9*]*

Prof. Yoichi Tanabe
Department of Applied Science, Okayama University of Science, Okayama, 700-0005, Japan
E-mail: tanabe@das.ous.ac.jp

Prof. Yoshikazu Ito
Institute of Applied Physics, Graduate School of Pure and Applied Sciences, University of Tsukuba, Tsukuba, Ibaraki 305-8573, Japan.
E-mail: ito.yoshikazu.ga@u.tsukuba.ac.jp

Prof. Katsuaki Sugawara,
Department of Physics, Graduate School of Science, Tohoku University, Sendai 980-8578, Japan

Prof. Katsuaki Sugawara, Prof. Takashi Takahashi, Prof. Mingwei Chen
Advanced Institute for Materials Research, Tohoku University, Sendai 980-8577, Japan

Prof. Mikito Koshino
Department of Physics, Osaka University, Osaka 560-0043 , Japan

Prof. Shojiro Kimura
Institute for Materials Research, Tohoku University, Katahira 2-1-1, Sendai 980-8577, Japan

Mr. Tomoya Naito
Department of Physics, Graduate School of Science, The University of Tokyo, Tokyo 113-0033, Japan
RIKEN Nishina Center, Wako 351-0198, Japan

Mr. Isaac Johnson, Prof. Mingwei Chen
Department of Materials Science and Engineering, Hopkins Extreme Materials Institute, Johns Hopkins University, Baltimore, MD 21214, USA
E-mail: mwchen@jhu.edu

Graphene is a two dimensional (2D) atomic-layer material with excellent electrical, chemical, thermal and mechanical properties for a wide range of applications [1–12]. However, practical applications of graphene are often limited by the intrinsic 2D morphology which only provides very limited areal mass loading and thus insufficient areal device performances. To amplify the device performances of 2D graphene, one approach is to construct three-dimensional (3D) graphene nano-architectures (denoted as 3D graphene). Various 3D graphene materials have been developed in the past decade [13–21]. However, it remains challenging to preserve the 2D graphene characters, such as Dirac fermions and 2D electron transport properties, in 3D graphene because of the introduction of crystal defects, loss of crystallinity, re-stacking of graphene sheets and high contact potentials between discrete graphene components. Recently, we have successfully fabricated 3D graphene by nanoporous metal based chemical vapor deposition. The resulting 3D graphene is constructed by a single graphene sheet with a bicontinuous and open porous structures (denoted as graphene sponge) and can fully preserve the 2D electronic properties of graphene with a large accessible specific surface area and high areal mass loading for a wide range of functional and structural applications [22–26].

One of the intriguing subjects of 3D graphene is the transport properties, which are an indispensable factor for electronic device applications. Typical structural components of the graphene sponge are 3D curved surfaces with possible grain boundaries, topological defects, and quasi-periodic structures. Theoretically, 3D curved surfaces give additional phases to conduction electrons through pseudo-magnetic fields [27] and, especially, a closed-loop path around topological defects which can flip the valley and pseudospin of an electron [28]. As a result, intravalley and an intervalley electron scattering could be enhanced when an electron propagates on 3D curved surfaces [27–29]. The formation of an energy gap, inversely proportional to the period of the 3D curved structure [30], has also been predicted from 3D periodic graphene [31–33]. Experimentally, the graphene sponge with a pore size (i.e., equivalent to curvature radius) of hundreds of nanometers shows graphene-like electric and optical properties [22, 34, 35].

Nearly 1000 times larger capacitance [34] and extremely low transmission in the optical response [35] have been observed as a consequence of the increase in the areal mass loading through 3D integration. However, since the curvature radii of these graphene sponge are 3–5 orders of magnitude larger than the interatomic distance of graphene, the geometric effect of curvature on electronic properties is anticipated to be insignificant, and the nonlinear magnetoresistance and Hall effects [34, 36] as well as broadband Dirac plasmonic absorptions [35] could be the consequence of topological and geometric effects [22, 37]. It is highly desired to fabricate high-crystallinity 3D monolayer graphene with a small curvature radius in the range from several to tens of nanometers by which the curvature effects on electronic behavior of graphene can be explored experimentally.

In this report, we systematically investigated the electronic states and electrical properties of graphene sponge with the curvature radius ranging from 25 – 1000 nm based on our recent success in fabricating high-quality 3D graphene using a nanoporous-metal-based chemical vapor deposition (CVD). This study provides experimental insights into the properties of 2D electrons on 3D curved surfaces of graphene and has important implications in designing and creating 3D graphene-based electronic devices.

The graphene sponges grown under different CVD conditions [22, 23] show 25 – 50 nm, 50 – 150 nm, and 500 – 1000 nm curvature radius, imaged by scanning electron microscopy (SEM) (**Figure 1a−c** and **Figure S1, S2**, Supporting Information). Although the substrates and CVD conditions are different, these samples have nearly identical bicontinuous nanoporous morphology in both geometry and topology except the difference in feature length. The average pore size and specific surface areas of the graphene sponges were quantitatively measured by the nitrogen adsorption/desorption method with Barrett-Joyner-Halenda (BJH) [38] and Brunauer–Emmett–Teller (BET) models [39]. The BJH pore sizes are well consistent with the SEM images (**Figure S3**, Supporting Information), and the BET surface areas range from 758.9 to 1260 $m^2/g$ [40, 41]. The low magnification TEM image (**Figure 1d**) of the sample with a 25 – 50 nm curvature radius (abbreviated as 25 – 50 nm sample) presents a bicontinuous porous structure that is comprised of a smoothly interconnected graphene sheet. The corresponding selected area electron diffraction pattern, as the inset of **Figure 1d,** shows sharp diffraction spots, verifying the high crystallinity of the graphene. The high-resolution TEM image (HRTEM) of the graphene (**Figure 1e**) taken from a curved region reveals the existence of topological defects, such as 5 and/or 7 membered rings (yellow circles), with local lattice distortions to accommodate the highly curved graphene lattices (red line).

Raman spectra reveal the relationship between the curvature radius and defect density. The 25 – 50 nm sample gives rise to a higher intensity of D bands (**Figure 1f** and **Table S1**, Supporting Information) than the graphene sponges with larger pore sizes, suggesting that a higher curvature leads to more geometrically-required defects. According to the intensity ratio of 2D and G bands ($I_{2D}/I_G$: 4.6 – 4.7), the graphene sponges are mainly comprised of high-quality monolayer graphene. The average distance between topological defects on the graphene lattice was estimated by the ratio of D and G bands in the Raman intensity spectra ($I_D/I_G$) with Tuinstra-Koenig relation [42–44] of $I_D/I_G = C(\lambda)/L_D$, where $C(\lambda)$ is a proportionality constant at the excitation laser wavelength $\lambda$ and $L_D$ is the average distance between the defects. The $L_D$ values of 25 – 50 nm sample ($I_D/I_G = 0.52$) is approximately 10 – 20 nm whereas the $L_D$ values of 50 – 150 nm ($I_D/I_G = 0.20$) and 500 – 1000 nm ($I_D/I_G = 0.04$) samples are about 20 – 50 nm and over 100 nm, respectively. It is worth noting that the difference in the defect density between the samples with the smallest and largest pores is only several times while the pore size (curvature radius) difference is 1 to 2 orders of magnitude. Therefore, out-of-plane elastic strains of graphene lattices may carry most of the curvature, while the topological defects may only play a minor role in forming 3D curved graphene. As a result, the obvious enhancement of D' band in the 25 – 50 nm sample is mainly contributed by intraband resonant Raman scattering from the 3D curved surface compared to larger pore size samples (**Table S1**, Supporting Information) and 2D graphene on a Cu sheet [5, 45]. Moreover, the binding state and quantitative chemical composition of the graphene sponges were investigated by X-ray photoelectron spectroscopy (XPS) (**Figure S4**, Supporting Information).

To elucidate the curvature dependence of electronic states of the graphene sponges, we performed photoemission spectroscopy (PES) measurements of the 3D graphenes with 25 – 50 nm, 50 – 150 nm (abbreviated as 50 – 150 nm sample), and 500 – 1000 nm (abbreviated as 500 – 1000 nm sample) in curvature radius (Supporting Infromation Section 2.3) using a customer-designed PES system (MBS A−1 spectrometer, MB Scientific AB) with a high-flux helium discharge lamp and a toroidal grating monochromator (MB Scientific AB) [46]. Graphene sponge samples were set at parallel to a photoelectron analyzer and the incident angle of photons with respect to sample normal is 45°. The energy and angular resolutions are 32 meV and ± 0.1°, respectively. As different orientations of graphene planes in the graphene sponges are exposed to the incident photons due to the unique structural nature of the graphene samples [22], the photoelectrons are emitted at various angles with respect to the graphene planes and thus the emission angle (momentum) dependence of photoelectrons is smeared out. As a result, only angle-integrated PES spectra can be obtained for investigating the electronic structure, especially valence bands, of the 3D curved graphene (**Figure 2a**). While the valence band structures in the 1000 nm pore sample have several features and resemble that of HOPG [47], the peaks around 4 – 9 eV of the 25 – 50 nm sample have much richer structures which are probably from $\pi+\sigma$ orbitals of highly curved graphene lattices. Compared to the 50 – 150 nm sample, the peak around 3 eV corresponding to $\pi$ orbitals becomes much weaker in the 25 – 50 nm sample, which also indicates the increase in the local interlayer interactions, the $sp^3$ configurations, and topological defects from the highly curved graphene lattices by reducing the curvature radii. Surprisingly, the DOS near $E_F$ in all samples keep the linear-like character of the 2D Dirac Fermions (inset of **Figure 2a**). However, the slopes of the linear DOS gradually decrease with curvature radii. Therefore, the Dirac Fermion system can be well preserved in the 3D curved graphene regardless of the curvatures, and importantly, it can be tuned by the highly curved surfaces.

The electrical transport properties of the graphene sponges were measured using electric double layer transistor (EDLT) devices to investigate the curvature-dependent electronic states. **Figure 2b** shows a gate voltage dependence of electrical resistance for the graphene sponges with various curvature radii (**Figure S5**, Supporting Information). The electrical resistances show ultralow values (0.8 – 2.2 Ω for the 500 – 1000 nm sample, 1.5 – 6 Ω for the 50 – 150 nm sample, 16 – 33 Ω for the 25 – 50 nm sample), which are ten to thousand times lower than that of 2D CVD graphene (500 – 15000 Ω) [5, 48] (**Table 1**). All samples show an electrical resistance peak in the $V_G$ range from −0.5 to 0.5 V, demonstrating the ambipolar electronic states of the graphene sponges [34]. The resistance on/off switching ratio is comparable between the graphene sponge samples (on/off ratio: 1.3 – 5) and much lower than those of 2D graphene EDLT (on/off ratio: ca.10) [49, 50], indicating no obvious bandgap formation across the entire graphene sponge samples. The systematic increase in the electrical resistance with the curvature radii decrease indicates that the curvature radius could be an essential parameter to tune the electronic properties of graphene sponges.

The dependence of the electrical resistance on the curvature radius was systematically investigated at low temperatures under magnetic fields. **Figure 3a** shows the temperature ($T$) dependence of the electrical resistance at zero magnetic fields for the samples with various curvature radii. The electrical resistance values progressively increase as the curvature radius decreases. The resistance curves of all the samples show a negative slope in the temperature ranges from 2 K to 300 K. **Figure 3b,c** are the zoom-in plots from the low-temperature region to elucidate the temperature dependence of the electrical conductance ($G$). The temperature-conductance curves obeys a logarithmic law at low-$T$, which is prominent for the 25 – 50 nm sample. Magnetic field ($B$) dependence of magnetoresistance was investigated, as shown in **Figure 3d−f**. For the 500 – 1000 nm sample, positive magnetoresistance is almost linear against *B* at 300 K. When the temperature decreases, the positive linear magnetoresistance is gradually suppressed. Below 10 K, negative magnetoresistance cusp develops at a low-*B* region, as shown in the inset of **Figure 3d**. For the 50 – 150 nm sample, the positive linear magnetoresistance amplitudes are reduced and the low-*B* cusps are enlarged at low temperatures. Interestingly, the 25 – 50 nm sample exhibits a distinctive magnetic field dependence. The positive linear magnetoresistance curves in the 3D graphene with a large curvature radius are completely replaced by the negative ones for all tested temperatures. At 1.9 K, a broadened negative magnetoresistance cusp is developed at low-*B*, and it smoothly connects to the shallow convex curve at high-*B*.

According to the tight-binding model of curved graphene, the overlap of π electrons outside the curved surface decreases but inside the curved surface increases (Supporting Information Section 3). As a result, the electron density tends to increase outside the curved surface due to Coulomb repulsion, and the transfer integral *t* (i.e., hopping between A and B sublattices of graphene) decreases with curvature radii. In this case, the effective mass $m^*$ is expected to increase with the decrease of the curvature radii. Experimentally, the linear electron density of states in the PES spectra of the graphene sponges demonstrate that the electronic state of 2D graphene is well preserved in the 3D curved nanostructure, which is consistent with the electrical transport measurements. While the electrical resistance, being proportional to the effective mass $m^*$, increases with the decrease of curvature radii, the gradient of PES spectra near the $E_F$, which is also associated with effective mass $m^*$, systematically decreases with curvature radii. Therefore, the inconsistence indicates that the geometric effect of 3D curvature may not be simply expected by the tight-binding model in the curved graphene. Alternatively, the local quasiperiodic potential, the circular pipe structure, and the high curvature induced pseudo magnetic fields may lead to the Bragg reflections of electron waves (i.e., energy gap opens locally) [30]. In this case, PES spectra could become a sum of massless and massive Dirac fermions and the changes of the DOS near $E_F$ depend on their volume fractions. Consequently, while the graphene sponges well preserves the electronic properties of 2D graphene, the 3D curved surface provides an additional degree of freedom to manipulate the 2D electron behavior and hence electronic properties.

In 2D disordered electron systems, diffusive electron motions from dominant elastic scattering may cause the interference among electron propagation pathways and results in the weak localization (WL) [51, 52]. The WL adds a logarithmic-law in the temperature-conductivity curves but it is difficult to be distinguished from electron-electron interactions (EEIs) [53]. In order to identify the WL effects, the magnetoconductivity measurement is essential. In 2D graphene, the intervalley backward scattering from charged impurities, edges, grain boundaries, etc. is the primary source of the WL corrections [32, 54, 55]. In the graphene sponges, 3D curved surfaces, together with grain boundaries and topological defects, could be the sources of intravalley electron scattering and/or intervalley electron scatterings [27−29, 36, 37, 56, 57]. The WL could be restored by the intervalley scattering through the miximg of *K* and *K'* valleys in the graphene electronic states. In fact, the number of possible scattering sources on the electron pathways increases when the curvature radius decreases. The negative magnetoresistance cusp in the low-*B* region, as a consequence of dephasing of WL states, gradually develops with the decrease of curvature radii and changes to the broadened negative one at the curvature radius of 25 – 50 nm. Note that the magnetoresistance at high-*B*, i.e., positive curves for the 50 – 1000 nm samples and shallow convex negative curve for the 25 – 50 nm sample, will be discussed in the next section.

To analyze the WL in the magnetoconductance where the low-*B* cusp is developed, we employed two models to fit the experimental data. The model I, formulated in **Equation (1)**, only considers the back-scattering effect [58] and the model II (**Equation (2)**) includes the graphene electronic structures [54, 55].

$$\Delta\sigma/\sigma_0 = \frac{e^2}{\pi h \sigma_0}\left[F\left(\frac{B}{B_\phi}\right) - F\left(\frac{B}{B_\phi + B_i}\right)\right], (1)$$

$$\Delta\sigma/\sigma_0 = \frac{e^2}{\pi h \sigma_0}\left[F\left(\frac{B}{B_\phi}\right) - F\left(\frac{B}{B_\phi + 2B_i}\right) - 2F\left(\frac{B}{B_\phi + B_i + B_*}\right)\right]. \quad (2)$$

In the equations above, $F(z) = \ln z + \psi(0.5 + z^{-1})$, $\psi(x)$ is a digamma function. $B_\phi, B_i, B_*$ are proportional to the phase breaking rate $\tau_\phi^{-1}$, the intervalley scattering rate $\tau_i^{-1}$, the intravalley scattering rate $\tau_*^{-1}$, respectively. The low-$B$ cusps in the magnetoconductance curves are fitted using **Equation (1)** and **(2)**, as shown in **Figures 4a−c**. The fitting ranges are limited to low-$B$, where the low-$B$ cusps in the magnetoconductance are developed. The magnetoresistance in high-$B$, which has different origins, will be discussed in the next section. Fitting parameters for **Equation (1)** and **(2)** at the lowest temperatures are displayed in **Figure S6** in the supporting infromation. The obvious increase in the scattering rates with decreasing of the curvature radius is confirmed by both models (**Figures 4a–c** and **Figure S6**, Supporting Information). The fact that the excellent fittings of the low-$B$ data from the three samples can be achieved from both models also indicates that intervalley scattering is essential, leading to the weak localization in the magnetoconductance. It is worth noting that large errors in fitting results of **Equation (2)** are due to the limited fitting range since the effect of weak antilocalization (WAL) (i.e., negative magnetoconductance curve) is involved in **Equation (2)**. For the fitting parameters of **Equation (2)**, the $B_i$ and the $B_*$ increases nearly 10 − 100 times when the curvature radius decreases to 25 – 50 nm. It suggests significant enhancements of intervalley and intravalley electron scattering from the geometric effect of the 3D curvature, which is consistent with the development of D and D' band in the Raman scattering. For quantitative comparison, $L_i = \hbar/4eB_i$ is in the range of 91 – 100 nm for 500 – 1000 nm sample, 40 – 47 nm for 50 – 150 nm sample, 14 – 16 nm for 25 – 50 nm sample, being approximate to the $L_D$ measured by the Raman spectra. Consequently, the WL correction in the electrical conductance is controlled by 3D curvature of the graphene sponges.

The high-$B$ magnetoresistance is dominated by the positive linear curves for the 500 – 1000 nm sample and decreases to be weakly positive for the 50 – 150 nm sample. The positive magnetoresistance in graphene is usually interpreted by WAL as a consequence of the $\pi$ Berry's phase of the massless Dirac fermion [54, 55]. However, the WAL corrections should be suppressed with the increase of temperatures, which is opposite to the experimental observations. To examine the WAL effects in the graphene sponges, temperature dependence of reduced electrical conductance ($\Delta G$), i.e. electrical conductance subtracting the electrical conductance at the lowest temperatures, was investigated for 50 – 150 nm samples under various magnetic fields (**Figure 4d, f**). A gradient of the log$T$ curve observed at zero magnetic fields is once reduced at 0.5 T and is enhanced from 3 T to 10 T. These suppression and enhancement of log$T$ curves at low temperatures demonstrates the dephasing of WL and WAL corrections by magnetic fields. The ratio of WAL to the positive linear magnetoresistance are deduced 5 – 10 % from gradients of log$T$ curves. Therefore, the present positive linear magnetoresistance mainly originates from semiclassical transport. In the 3D curved graphene structure, the effective transverse magnetic field locally weakens depending on the angle between the 3D curved surface and the magnetic fields, which can result in the positive linear magnetoresistance from the semiclassical model [34, 36] (Supporting Information Section 4.1, 4.2). In this scenario, the suppression of the positive linear magnetoresistance observed from 50 – 150 nm sample, in comparison with that from 500 – 1000 nm can be explained by the decreased carrier mobility from the additional electron scattering caused by the 3D curved surfaces. For the 25 – 50 nm sample, the positive magnetoresistance, originating from the semiclassical transport and the WAL corrections, is completely suppressed by the electron scattering effects from 3D curvature and only the shallow convex negative magnetoresistance is left, together with the WL correction at low-$B$. This broad negative magnetoresistance at high-$B$ can be attributed to EEIs. When the curvature radius further decreases to 25 – 50 nm, highly curved 3D surface significantly enhances both intravalley and intervalley electron scattering ($B_i/B_\phi, B_*/B_\phi > 1$ ; $B_i, B_*, B_\phi \propto \tau_i^{-1}, \tau^{*-1}, \tau_\phi^{-1}$) . Such diffusive electron motion could amplify the EEIs between scattered electrons [52, 59–62]. The magnetoresistance subtracting the WL correction as shown in **Figure 4c** using **Equation (1)** and **(2)** (Open circle and square in **Figure 4g**) is proportional to $-B^2$ curves (red lines) [52, 62, 63]. This quadratic curve is the characteristic behavior of the EEI correction . To elucidate a component of the EEI corrections in the temperature-conductance curves, which are well known as $\log T$ curves, the temperature dependence of electrical conductance at high-$B$ was investigated because the $\log T$ correction originating from EEI is insensitive to the magnetic fields in the present conditions, in contrast to the WL correction. When WL corrections are well suppressed above 5 T, the electrical conductance proportional to $\log T$ is represented up to 15 T. The gradients of $\log T$ curves are almost constant as shown in **Figure 4e, f**, consistent with the EEI corrections. The saturation of the gradients of $\log T$ curves above 10 T for 50 – 150 nm samples as shown in **Figure 4d, f** can also be interpreted as EEI corrections. For 25 – 50 nm samples, the gradients of $\log T$ curves are obviously larger than those of 50 – 150 nm samples, indicating the enhancement of the EEI corrections by the electron scattering of the 3D curved surface. Therefore, the highly curved 3D surface significantly affects the transport properties of 2D graphene by suppressing the $\pi$ Berry's phase transport and enhancing the electron-electron interactions (Supporting Informations Section 4.5).

We make a comparison of electric performances of graphene sponges with 2D graphene, as shown in **Table 1**, for understanding the curvature effects in graphene properties. Comparing with 2D graphene with electrical conductance of $10^{-4}$ – $10^{-3}$ S [5, 47, 63, 64] and capacitance of 0 – 0.005 mFcm$^{-2}$ [49], the 500 – 1000 nm graphene sponge can be simplified as the parallel circuits of a single graphene sheet since the electrical capacitance of the graphene sponge is ca.1000 times higher than that of 2D graphene [34] (**Figure S7**, Supporting Information). Under this assumption, it can be understood that the suppression of the electrical conductance in the 25 – 50 nm and 50 – 150 nm samples comes from additional scattering effects originating from the geometry of graphene sponges (**Figure S8**, Supporting

Information). Thus, while the capacitance of the graphene sponges can be enhanced by the increase of the integration efficiency in the graphene nano-architectures, the electrical conductance could be limited by the electron scattering effects from 3D curvature.

The Hall mobility for 500 – 1000 nm sample is calculated from normalized electrical conductance. The Hall coefficient at maximum magnetic fields is ca. 600 – 700 cm$^2$V$^{-1}$s$^{-1}$, which is comparable with the lower bound of 2D – CVD graphene. The value decreases with the curvature radius, as shown in **Table 1**. However, the conventional Hall mobility may underestimate the carrier mobility because 3D electron pathways can weaken the actual transverse magnetic field from place to place [34, 36]. To estimate the actual carrier mobility, we previously proposed a simple model based on the simplified 3D graphene structures [34, 36] (Supporting Information Section 4.1, 4.2), although a more accurate model would require a sophisticated method [65]. Carrier mobility of the 25 – 50 nm sample is 10−100 times lower than that of the 500 – 1000 nm sample , which is quantitatively in agreement with the enhancement of $B_i, B_*$, and underline the additional effects in electron scattering and carrier mobility from the 3D curved structure.

We have elucidated the effects of 3D curvatures on the electronic states and electrical transport properties of graphene. It has been found that the electronic states of 2D graphene, i.e., linear density of states and ambipolar electronic transport properties, can be well preserved in 3D curved graphene with the curvature radius down to 25 – 50 nm. The 3D curvature can effectively suppress the electron density of states near the Fermi energy, providing a new degree of freedom to tune the electronic properties of graphene. Moreover, the curvature enhances the scattering of 2D Dirac fermion through the electron scattering (interference) in the 3D curved space. As a result, the conductivity corrections of both the electron localization and electron-electron interaction suppress the electrical transport in 3D curved graphene. This study experimentally unveils the effect of 3D curvature on electronic states and electrical transport properties of graphene, providing new insights into the 2D electron kinetics in 3D nanoarchitecture with intrinsic curvature and essential knowledge to amplify various excellent properties of graphene for 3D device applications.

**Acknowledgments**

We thank Kazuyo Omura at the Institute for Material Research in Tohoku University for XPS measurements. This work was sponsored by JSPS Grant-in-Aid for Scientific Research on Innovative Areas "Discrete Geometric Analysis for Materials Design": Grant Number JP18H04477, JP20H04628; JSPS KAKENHI Grant Number JP15H05473, JP18K14174, JP17K14074, JP18K18986, JP18H01821, JP19K05195, JP19J20543; World Premier International Research Center Initiative (WPI), MEXT, Japan; NIMS microstructural characterization platform as a program of "Nanotechnology Platform Project", MEXT, Japan; Izumi Science and Technology Foundation; a cooperative program (Proposal No. 20G0002) of the CRDAM-IMR, Tohoku University. This work was performed at High Field Laboratory for Superconducting Materials, Institute for Materials Research, Tohoku University (Project No. 18H0205, 19H0204, 20H0011), the RIKEN iTHEMS program. M.C. was sponsored by the Whiting School of Engineering, Johns Hopkins University and National Science Foundation (NSF PD 09-1771).

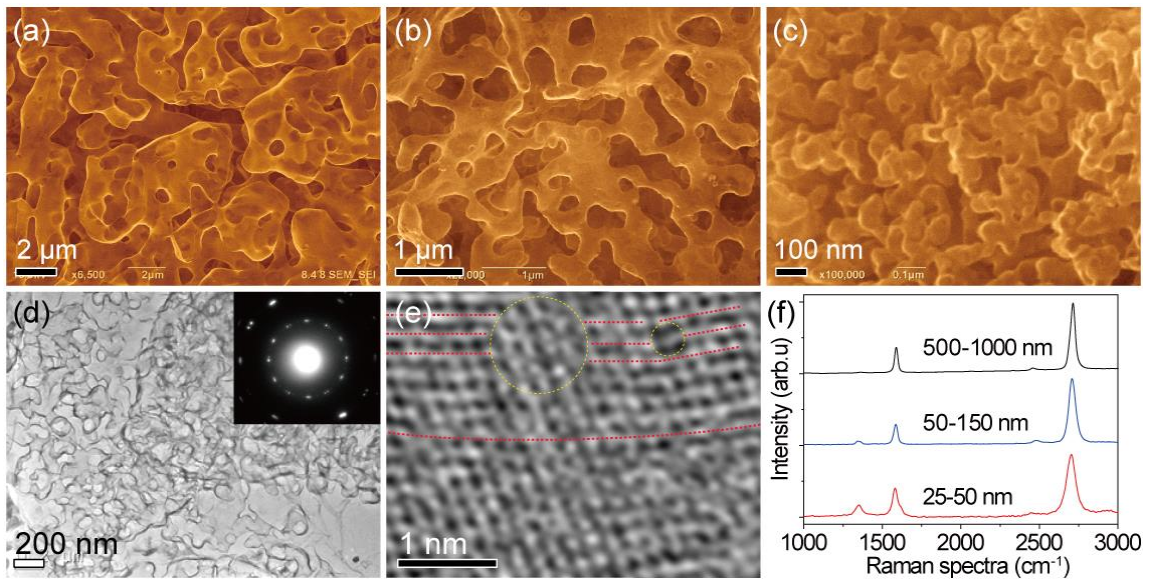

**Figure 1.** Scanning electron microscope images of graphene sponges with the curvature radii of **a)** 500 –1000 nm, **b)** 50 – 150 nm and **c)** 25 – 50 nm. **d)** Low magnification TEM images of 25 – 50 nm sample. Inset of **d)** shows the corresponding selected area electron diffraction patterns. **e)** HRTEM image of **d)** taken from a curved region. Yellow circles and red lines present topological defects such as 5 and/or 7 membered rings and visible lattice distortion of the carbon hexagon matrix. **f)** Raman spectra of the graphene sponge samples.

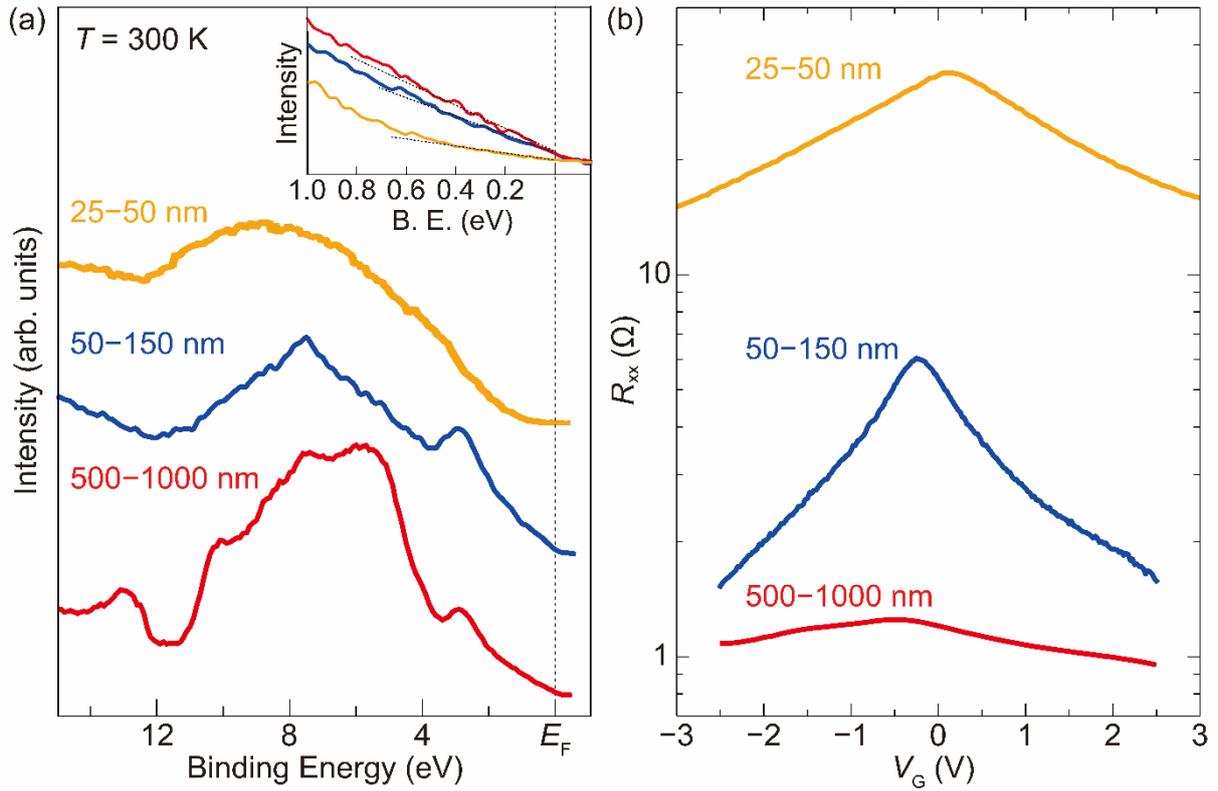

**Figure 2.** Angle-integrated photoemission spectra (PES) of graphene sponges measured by He IIa resonance line ($hv = 40.814$ eV) at room temperature. **a)** 25 – 50 nm (orange), 50 – 150 nm (blue), and 500 – 1000 nm (red) curvature radii samples. The inset shows the angle-integrated PES near the Fermi level. The black dash lines are the linear extrapolations to the Fermi level. **b)** Longitudinal resistance against the gate voltage $V_G$ for the graphene sponge EDLT with 25 – 1000 nm curvature radii at room temperature.

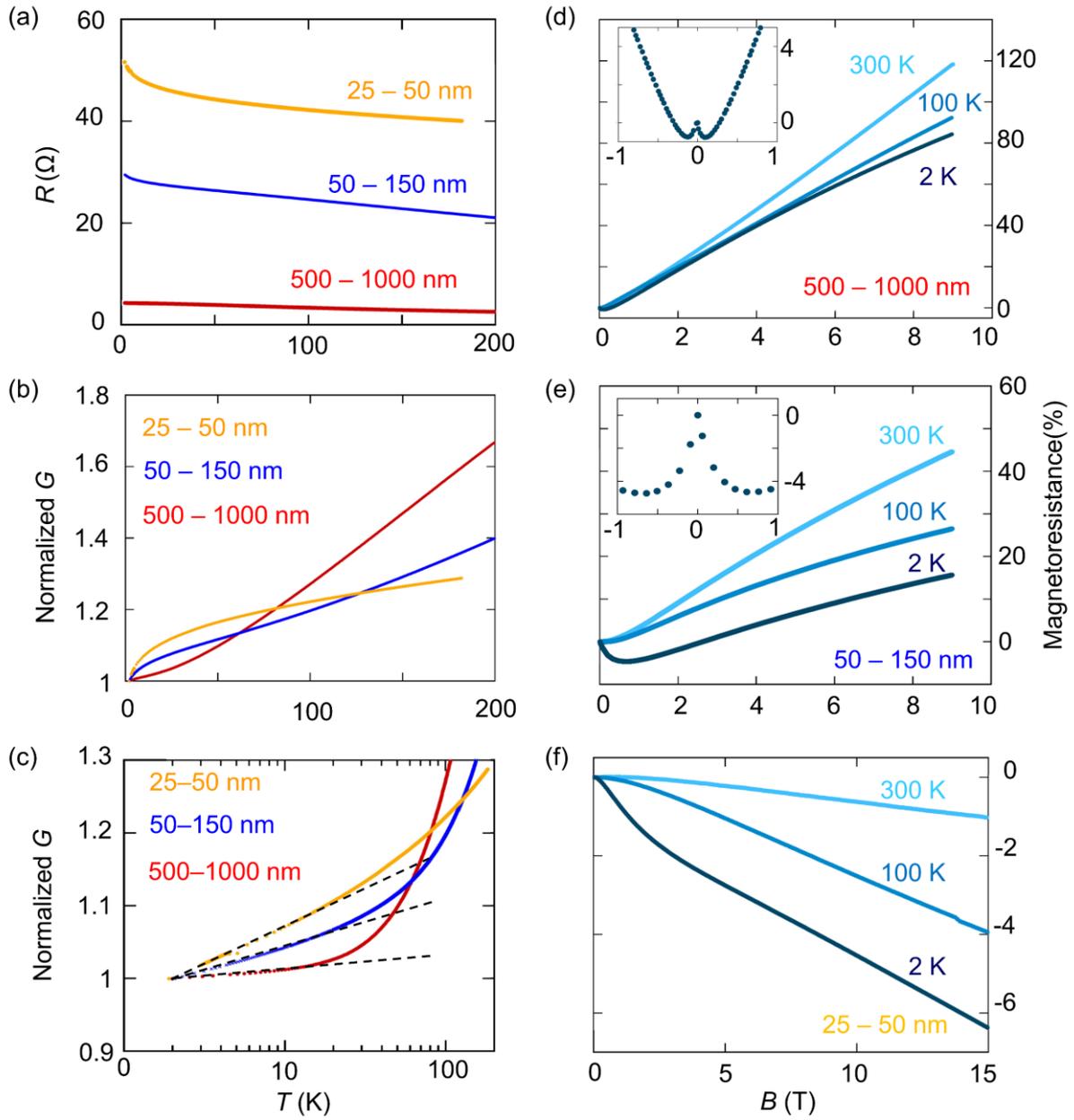

**Figure 3.** Electrical transport of graphene sponges with 25 – 1000 nm curvature radii. **a)** Temperature-resistance curves. **b, c)** Temperature-conductance curves. **d-f)** Magnetoresistance curves. Insets of **d,e)** are the magnified plots at low-*B* regimes.

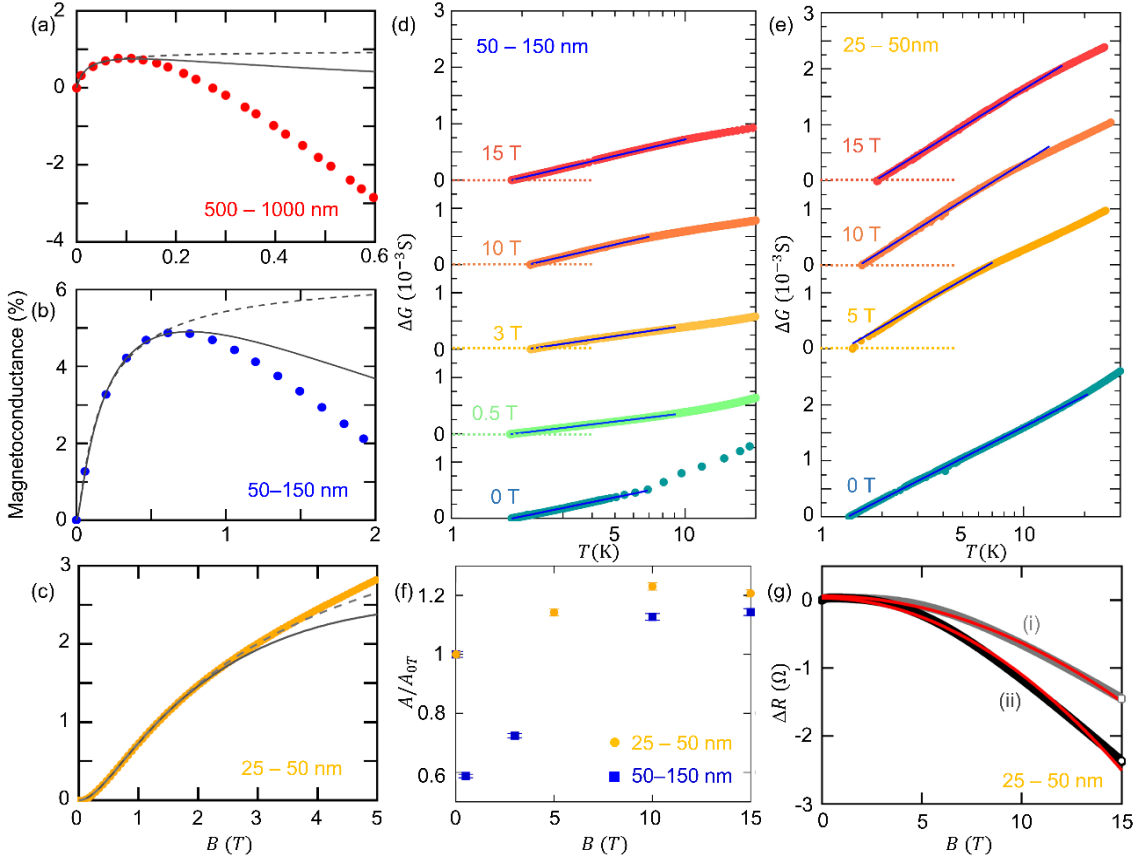

**Figure 4. a-c)** Magnetoconductance curves of graphene sponges with 25 – 1000 nm curvature radii. Broken and solid black lines are the fitting results of **Equation (1)** and **(2)** to extract the weak localization corrections from the magnetoconductance curves. $B_\phi, B_i, B_*$ are the fitting parameters that are proportional to the phase breaking rate $\tau_\phi^{-1}$, the intervalley scattering rate $\tau_i^{-1}$, and the intravalley scattering rate $\tau_*^{-1}$. **d, e)** Temperature-reduced conductance ($\Delta G$) curves for 50 – 150nm and 25 – 50 nm under various magnetic fields. $\Delta G$ are calculated by subtracting the conductance at the lowest temperatures from raw conductance curves. Blue lines present the fitting results using $\Delta G = A\log T + B$. **f.)** Magnetic fields dependence of gradients of logarithmical temperature conductance curves $A$, normalized by $A$ at 0 T ($A_{0T}$). Error bars are estimated from the results of the least squared fitting. For 50 – 150 nm sample, the logarithmical temperature-reduced conductance curves are first suppressed around 0.5 T due to weak localization and then enhanced from 0.5 T to 10 T from weak antilocalization. The saturation of the normalized $A$ at around 10 – 15 T is due to the electron-electron interactions (EEIs). For 25 – 50 nm, the normalized $A$ is almost constant around 5 – 15 T, originating from the EEIs. **g)** The magnetic field dependence of resistance after subtracting the WL corrections in **Equation (1)** and **(2)** (closed circle (black) and square (gray)) for 25 – 50 nm sample. Red lines present the fitting results using quadratic curves.

**Table 1.** Electric performances of graphene sponges. Electrical conductances are normalized by the geometrical dimension of graphene sponge.

|  | 25 − 50 nm graphene sponge | 50 − 150 nm graphene sponge | 500 − 1000 nm graphene sponge | CVD graphene |
|---|---|---|---|---|
| Electrical Conductance at 300 K (S) [a] | 0.03−0.05 | 0.06−0.5 | 0.5−1 | ca.$10^{-4}$ − ca.$10^{-3}$ [5,48] |
| Capacitance (mFcm$^{-2}$) | 2.0−3.6 | 1.0−2.5 | 0.40−1.0 | 0-0.005 [c] [49] |
| Carrier mobility [b] (cm$^2$V$^{-1}$s$^{-1}$) | 7−8 | 100−200 | 600−700 | 550 − 25000 [5, 48, 63, 64] |
| Error bar | 4 | 100 | 400 |  |
| Carrier mobility in eq. (SE13) [b] (cm$^2$V$^{-1}$s$^{-1}$) | ca. 1000 | ca. 8000 | ca. 20000 | N/A [d] |
| Error bar | 500 | ca. 4000 | ca. 5000 |  |

[a] Sample dimensions for normalized electrical conductance are 1 mm channel length, 1 mm width, and 0.030 mm thickness.

[b] Carrier mobilities of the graphene sponges were estimated from Hall mobility in the single carrier model and **Equation (SE13)** in the supporting information. Hall mobility is calculated using the electrical conductance and Hall coefficient defined at maximum magnetic fields. The second derivative of **Equation (SE13)** in the supporting information shows the peak at $\mu B = 0.7$, which estimates the carrier mobility.

[c] Results in the exfoliated graphene.

[d] Mobility in **Equation (SE13)** in the supporting information for 2D-CVD graphene cannot be estimated since Hall resistivity is linear against magnetic fields. (Supporting Information Section 4.4)

**Dirac Fermion kinetics in three-dimensionally curved graphene**

By *Yoichi Tanabe[1*], Yoshikazu Ito[2*], Katsuaki Sugawara[3,4], Mikito Koshino[5], Shojiro Kimura[6], Tomoya Naito[7,8], Isaac Johnson[9], Takashi Takahashi[4], Mingwei Chen[4,9*]*

Prof. Yoichi Tanabe
Department of Applied Science, Okayama University of Science, Okayama, 700-0005, Japan
E-mail: tanabe@das.ous.ac.jp

Prof. Yoshikazu Ito
Institute of Applied Physics, Graduate School of Pure and Applied Sciences, University of Tsukuba, Tsukuba, Ibaraki 305-8573, Japan.
E-mail: ito.yoshikazu.ga@u.tsukuba.ac.jp

Prof. Katsuaki Sugawara,
Department of Physics, Graduate School of Science, Tohoku University, Sendai 980-8578, Japan

Prof. Katsuaki Sugawara, Prof. Takashi Takahashi, Prof. Mingwei Chen
Advanced Institute for Materials Research, Tohoku University, Sendai 980-8577, Japan

Prof. Mikito Koshino
Department of Physics, Osaka University, Osaka 560-0043 , Japan

Prof. Shojiro Kimura
Institute for Materials Research, Tohoku University, Katahira 2-1-1, Sendai 980-8577, Japan

Mr. Tomoya Naito
Department of Physics, Graduate School of Science, The University of Tokyo, Tokyo 113-0033, Japan
RIKEN Nishina Center, Wako 351-0198, Japan

Mr. Isaac Johnson, Prof. Mingwei Chen
Department of Materials Science and Engineering, Johns Hopkins University, Baltimore, MD 21214, USA
E-mail: mwchen@jhu.edu

# 1. Materials and methods
## 1.1 Fabrication of nanoporous Ni substrates

$Ni_{30}Mn_{70}$ alloy ingots were prepared by melting pure Ni and Mn (purity >99.9 at.%) using an Ar-protected arc melting furnace. After annealing at 900°C for 24 hours for microstructure and composition homogenization, the ingots were cold-rolled to thin sheets with a thickness of ~50 μm. Nanoporous Ni was prepared by chemical dealloying in a 1.0 M $(NH_4)_2SO_4$ aqueous solution at 50°C overnight. After the dealloying, the samples were rinsed thoroughly with distilled water and then rinsed with ethanol before dried in vacuum.

## 1.2 Synthesis of graphene sponges by CVD

Nanoporous Ni substrates were loaded in an inner quartz tube (diameter 26 mm × diameter 22 mm × 250 mm length) which was inserted into a larger quartz tube (diameter 30 mm × diameter 27 mm × 1000 mm length) in an open-box-type furnace. The Ni substrates were annealed under flowing gas mixed with 200 sccm Ar and 100 sccm $H_2$, which is monitored by mass flow controllers. For the sample with 50−150 nm curvature radius, the Ni substrates were annealed at 800°C for 3 min. For the sample with 500−1000 nm curvature radius, the Ni substrate were annealed at 900°C for 57 min. After the reduction pre-treatment, the mixed gases of $H_2$ (100 sccm), Ar (200 sccm), and benzene, (0.5 mbar, 99.8%, anhydrous, Aldrich) as graphene precursors were introduced into the inner tube for graphene growth. The CVD growth was performed at 800°C for 30 seconds for the 50−150 nm sample, and at 900°C for 30 seconds for the 500−1000 nm sample. The growth proceeded with quick cooling of the inner and outer quartz tubes after CVD growth by opening the furnace. The nanoporous Ni substrates were dissolved by 1.0 M HCl solution overnight and the resulting graphene sponges were then transferred into a 2.0 M HCl solution to completely removed residual Ni. The samples were repeatedly washed with water and kept in IPA for a supercritical $CO_2$ fluid drying.

## 1.3 Preparation of NiO nanoparticles

NiO nanoparticles were fabricated by a standard hydrothermal synthesis, as shown in Scheme S1 [S1]. The stock solution of NiO nanoparticles was prepared with 1.0 M $Ni(NO_3)_2$ $6H_2O$ (99.9%, Wako) and two equivalence of KOH (99.8 %, Wako) as a pH adjustment [S2]. The NiO nanoparticles were synthesized with hydrothermal and supercritical water processes (MomiCho-mini, ITEC Co., Ltd.) by using the stock solution. The stock solutions were kept stirring with a magnetic stirrer in a slurry pump (MS-02-AD, ITEC Co., Ltd.) and reacted with pure water at 400°C under 30 MPa. The typical flow rate of stock solution and pure water was 3 ml/min and 10 ml/min, respectively. The typical reaction time was 1 second. The resulting solutions were collected by cylinders and pushed out to a 500 mL beaker. The resulting NiO nanoparticles were centrifuged and thoroughly washed by distilled water.

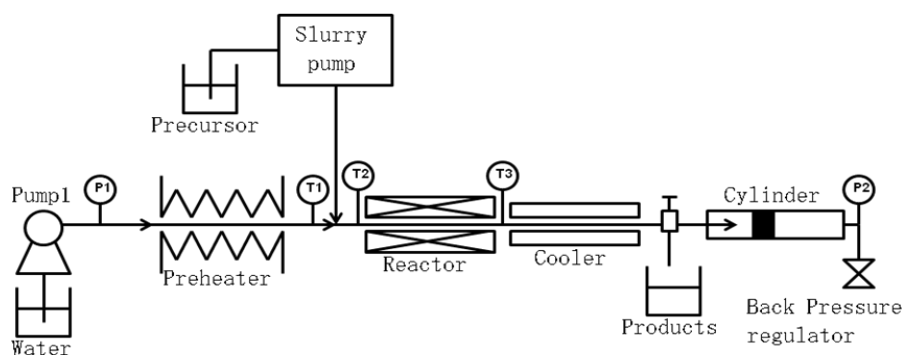

Scheme S1. Procedures for hydrothermal synthesis. P = pressure gauge and T = thermocouple.

### 1.4 Preparation of graphene sponges by NiO nanoparticle-based CVD method

The as-prepared NiO nanoparticles were dispersed in ethanol and cast on Cu sheets as CVD templates. The nanoparticles on the Cu sheets were loaded in the inner quartz tube (diameter 26 mm × diameter 22 mm × 250 mm length) which was inserted into the center of a larger quartz tube (diameter 30 mm × diameter 27 mm × 1000 mm length) in the open-box-type furnace. The nanoparticles were annealed at 350°C for 25 min to form nanoporous NiO by sintering and continuously annealed at 750°C for 2 min to reduce nanoporous NiO surface to metal Ni under a mixed atmosphere of $H_2$ (100 sccm) and Ar (200 sccm). Graphene was grown at 750°C for 30 seconds using a mixed gas of $H_2$ (100 sccm), Ar (200 sccm), and benzene (0.5 mbar, 99.8%, anhydrous, Ardrich) as graphene precursors. The nanoporous Ni substrates were dissolved by 1.0 M HCl solution overnight and then transferred into 2.0 M HCl solution, to dissolve residual Ni at 50°C. The samples were repeatedly rinsed with distilled water for five times and kept in IPA for a supercritical $CO_2$ fluid drying.

### 1.5 Procedure of supercritical drying

To prevent collapse and damage of the 3D curved structure of graphene sponges by the capillary force of water during drying, the graphene sponge samples were immersed in IPA and dried using supercritical $CO_2$ (sc$CO_2$) to gradually substitute IPA with minimized capillary force.[S3] The samples were first transferred to a glass bottle (volume: 5 mL) filled with IPA (400 μL), and then the bottle was placed in an 80 mL pressure-resistant container (TAIATSU techno Corp). After removing the air inside the container with gaseous $CO_2$ purging, the pressure of the container gradually increased to 15 MPa by introducing liquid $CO_2$ (5 MPa, −4°C, the density of 0.964 g/mL) at a flow rate of 20 mL/min (19 g/min) using a high-pressure plunger pump (NIHON SEIMITSU KAGAKU Co. Ltd, NP-KX-540). The sc$CO_2$ drying process was carried out at 70°C with constant $CO_2$ flowing at a rate of 5 mL/min (4.8 g/min) by forming a homogeneous phase of IPA and sc$CO_2$ to minimize the capillary force. The pressure was maintained at 15 MPa during the drying procedure for 5 h at least. After drying completely, the temperature was set at 40°C, and it was gradually depressurized over 43 h from 15 MPa to atmospheric pressure by gradually venting $CO_2$ from the system.

## 2. Microstructure characterization and property measurements
### 2.1 Microstructural characterization

The microstructure of graphene sponge samples was characterized by scanning electron microscope (SEM, JEOL JSM−6700) and transmission electron microscope (JEOL JEM−2100F) with two aberration correctors for the image- and probe-forming lenses. SEM observations were conducted at an accelerating voltage of 15.0 kV, and the samples were placed on conductive carbon tapes. For TEM observations, the samples were transferred on a Cu grid without a carbon support film. Chemical compositions of the graphene sponge samples were measured with electron energy-loss spectroscopy (EELS, JEOL) and X-ray photoelectron spectroscopy (XPS). Nanopore sizes and surface areas of the graphene sponge samples were measured with the Brunauer−Emmett−Teller (BET) and Barrett−Joyner−Hallender (BJH) models using a BELSORP-mini II (BEL. JAPAN. INC) at 77 K. The horizontal axis was normalized with the vapor pressure of nitrogen ($P_0$) at 77.0 K (= 0.101 MPa). All samples were heated at 120°C under vacuum for 48 hours before the measurements. The mass of the samples was measured with an ultramicro balance.

### 2.2 Raman spectroscopy of graphene sponges

Raman spectra were recorded by using a micro-Raman spectrometer (Renishaw InVia RM 1000) with an incident wavelength of 514.5 nm. The laser power was set at 2.0 mW to avoid possible sample damage by laser irradiation. The graphene sponge samples were placed on a background-free glass slide. The accumulation time of each spectrum is 400 s.

### 2.3 Photoelectron emission spectroscopy of graphene sponges

Photoemission spectroscopy (PES) measurements were performed using a MBS A−1 spectrometer (MB Scientific AB) with a high-flux helium discharge lamp and a toroidal grating monochromator (MB Scientific AB) at room temperature. He IIα resonance line ($hv$ = 40.814 eV) was employed for photoelectron excitation. The energy and angular resolutions were set at 32 meV and ± 0.1°, respectively. Graphene sponge samples were mounted on an n-type Si(111) wafer substrates with low electrical resistance (As doping, 1−5 mΩcm) to avoid the charging effect during PES measurements. Graphene sponge samples were set at parallel to the photoelectron analyzer and the incident angle of photons with respect to the sample normal was 45°. Before PES measurements, the graphene sponge samples were annealed at 600°C under ultrahigh vacuum at 1×10$^{-10}$ Torr for 3 hours. The Fermi levels of the samples were referred to that of a gold film deposited on the sample holder. Angle-integrated photoemission spectra were integrated over the various emission angle (momentum) of photoelectrons emitted from the crystal surfaces, which corresponded to the valence-band density of states in the samples. While the 2D graphene sheets show a strong emission-angle dependence of PES spectra due to an atomically flat surface, the photoemission spectra of the graphene sponge do not show the angle (momentum) dependence because the graphene sponge was constructed by randomly orientated curvature gradients, resulting in angle-integrated photoemission spectra [S3].

### 2.4 Measurements of electrical transport and capacitance using EDLT device

The gate electrodes in the EDLT devices were prepared with the graphene sponges [S4]. Two pieces of graphene sponges were pasted on a glass substrate using the double-faced tape. Both gate and channel electrodes were made by sputtering platinum. In order to prepare the electrical contact between device and measurement system, Cu polyimide wires were attached on the electrodes with silver paste. Pt electrodes and Ag paste were covered by a silicone resin. An ionic liquid, *N*, *N*-Diethyl-*N*-methyl-*N*-(2methoxyethyl)ammonium Bis(trifluoromethanesulfonyl)imide (DEME-TSFI), was employed as the electrolyte of the EDLT device, which was often used to obtain stable EDLT performances of carbon materials in a large gate voltage ($V_G$) range[S5]. The DEME-TSFI was draped on the entire device to form an electric double-layer capacitor between the transistor channel and the gate electrode. The ionic liquid was prebaked at 120 °C under the vacuum condition to remove water moisture.

The graphene sponge EDLT devices were measured using the standard four-probe method. $^4$He cryostat with a superconducting magnet was employed to control temperature and magnetic field. For the Hall resistance ($R_{yx}$) measurements, raw $R_{yx}$ was symmetrized to remove the effects of longitudinal resistance ($R_{xx}$) due to the misalignment of electrodes. The capacitance was measured using the potentiostat/galvanostat (IVIUM Technologies) under a constant voltage between the transistor channel and the gate voltage (10$^0$−10$^5$ Hz). To make a comparison of capacitance between 2D graphene and graphene sponge, the capacitance was normalized by the projected area of graphene sponge. The total capacitance was described by:

$$C = \left(\frac{1}{C_Q} + \frac{1}{C_L}\right)^{-1},$$

where $C_Q$ was the quantum capacitance, and $C_L$ was the geometrical capacitance, respectively. For the EDLT devices, a large geometrical capacitance was expected, but a quantum capacitance, being proportional to the finite density of states of graphene, was dominant in the total capacitance [S6−S8]. The capacitance of the transistor channel was calculated with the assumption that the surface area of the reference electrode was sufficiently large.

## 2.5 Measurements of low-temperature electrical transport

The graphene sponges were pasted on a glass substrate using the double-faced tape. Channel electrodes were made by sputtering platinum. In order to prepare the electrical contact between sample and measurement system, Cu polyimide wires were attached to the electrodes with silver paste. Low-temperature electrical transports of graphene sponges were measured using the standard four-probe method. $^4$He cryostat with a superconducting magnet was employed to control temperature and magnetic field. Cernox thermometer was employed to measure the temperature of the sample. We confirmed that temperature errors of Cernox thermometer are around 0.1 K in the present temperature and magnetic field ranges and do not affect current interpretations. For the Hall resistance measurements, obtained raw Hall resistance curves were symmetrized to remove the effects of longitudinal resistance due to the misalignment of electrodes.

## 3. Tight-binding model of graphene
### 3.1 Case of 2D graphene

The primitive cell of the 2D graphene consists of two carbon atoms A and B. The Hamiltonian of the 2D graphene in the tight-binding model is

$$\mathcal{H} = t\Sigma_{i,j,\sigma}(a^\dagger_{\sigma i}b_{\sigma j} + b^\dagger_{\sigma j}a_{\sigma i}), \text{(SE1)}$$

where $t \sim -2.8$ eV is the transfer integral, and $a$ and $b$ are the annihilation operators at carbon A and B, respectively. Here, only the nearest-neighbor hopping is considered. The eigenvalue of the Hamiltonian reads:

$$E_\pm(k) = \pm t\Sigma \sqrt{3 + 2\cos(\sqrt{3}k_y a) + 4\cos\left(\frac{3}{2}k_x a\right)\cos\left(\frac{\sqrt{3}}{2}k_y a\right)}, \text{(SE2)}$$

where $a$ is the bond length and $k = (k_x, k_y)$ is the wave number[S9]. The effective mass $m^*$ is, in general, defined by

$$m^* = \left(\frac{\partial^2 E(k)}{\partial k^2}\right)^{-1} \text{(SE3)}.$$

The relationship between the electrical resistance $R$ and the effective mass is

$$R = \frac{m^*}{e^2 n\tau} \text{(SE4)}$$

in the simple Drude model, where $n$ and $\tau$ denote the electron density and the relaxation time [S10].

In $\pi$ and $\pi^*$ orbitals of the 2D graphene, the effective mass $m^*$ is proportional to the inverse of the transfer integral $1/t$, and hence the electric resistance $R$ is also proportional to $1/t$.

The density of states is also affected by the transfer integral $t$ [S11]. As the transfer integral $t$ decreases, the density of state near the Fermi energy increases.

### 3.2 Case of 3D graphene

The 3D graphene with the curvature radius $r$ is considered. The $\pi$ electron of the carbon atom A at the interior of the curved surface becomes closer to that of the atom B. Due to the Coulomb repulsion between them, the density of π electrons at the interior of the curved surface decreases and, at the same time, the density of π electrons at the exterior of the curved surface increases. Thus, the contribution of π electrons at the interior of the curved surface to the transfer integral $t$ decreases due to the decrease of the density of π electrons. Moreover, the contribution of π electrons at the exterior of the curved surface also decreases due to the decrease of the overlap between π electrons. Thus, the transfer integral $t$ totally decreases as the curvature radius $r$ decreased.

Under the assumption that the 3D graphene keeps the electronic structure of the 2D graphene and only the transfer integral $t$ changes, both the electrical resistance $R$ and the density of state increases as the curvature radius $r$ decreases. In reality, the decrease of the relaxation time $\tau$ caused by the topological defects and the curved surfaces also increases the electrical resistance $R$.

## 4. Supporting discussion

### 4.1 Semiclassical theory for the electrical transport under magnetic field in graphene sponge

When the graphene sponge was placed in a uniform magnetic field, the relative angle between the magnetic field and the normal of the curved graphene surface varied from place to place. Since the electron orbital motion was only affected by the magnetic field component perpendicular to the plane, an electron traveling on the graphene labyrinth felt a non-uniform magnetic field, and thus the total conductivity should be expressed as an average of the 2D conductivity over the magnetic field amplitude when the inhomogeneity varied slowly. Here we presented a semiclassical picture to describe the magnetotransport along this line and showed that the approach well described the magnetic-field dependence of the Hall resistance and the magnetoresistance was observed in the experiments [S4, S12].

We first considered a flat two-dimensional electron system in a perpendicular magnetic field $B$. In the semiclassical approximation, the conductivity tensor was given as

$$\begin{pmatrix} \sigma_{xx}^{2D} & -\sigma_{xy}^{2D} \\ \sigma_{xy}^{2D} & \sigma_{xx}^{2D} \end{pmatrix} = \frac{\sigma_0^{2D}}{1+\omega_c^2\tau^2} \begin{pmatrix} 1 & -\omega_c\tau \\ \omega_c\tau & 1 \end{pmatrix}, \text{(SE5)}$$

where $\omega_c$ was the cyclotron frequency, $\tau$ was the scattering time, and $\sigma_0^{2D}$ was the conductivity in zero magnetic field. For graphene, the conductivity was simplified as [S13]

$$\sigma_0^{2D} = n_s e \mu, \text{(SE6)}$$

where $n_s$ was the carrier concentration and $\mu$ was the mobility given by

$$\mu = \frac{ev^2\tau}{\varepsilon_F}, \text{(SE7)}$$

here $v$ was the constant band velocity of graphene, and $\varepsilon_F$ was the Fermi energy. The cyclotron frequency was expressed as

$$\omega_c = \frac{ev^2 B}{\varepsilon_F}, \text{(SE8)}$$

So, we ended up with a relation

$$\omega_c \tau = \mu B, \text{(SE9)}$$

as in ordinary metals.

Now let us consider the conductivity of graphene sponge in a uniform magnetic field $B$ along the $z$-direction. Here we assumed that the normal of the local graphene surface was distributed isotropically and that the conductivity contributed by surface elements was additive, although a more accurate picture would require a sophisticated method such as the random network theory. The total three-dimensional conductivity tensor then became

$$\sigma_{ij}(B) = \frac{S_{tot}}{V} \int_0^{\frac{\pi}{2}} \sigma_{ij}^{2D}(B\cos\theta) \sin\theta \, d\theta, \text{(SE10)}$$

where $\theta$ was the angle from the surface normal to the magnetic field, $S_{tot}$ and $V$ were the total graphene area and the total volume of the system, respectively. The integral could be analytically evaluated to give

$$\sigma_{xx}(B) = \sigma_0 \frac{\tan^{-1}\omega_c\tau}{\omega_c\tau}, \quad \sigma_{xy}(B) = \sigma_0 \frac{\log(1+\omega_c^2\tau^2)}{2\omega_c\tau}, \text{(SE11)}$$

where $\sigma_0 = (S_{tot}/V)\sigma_0^{2D}$. The resistivity tensor was given by inverting the conductivity tensor as

$$\rho_{xx}(B) = \rho_0 \omega_c \tau \frac{\tan^{-1}\omega_c\tau}{(\tan^{-1}\omega_c\tau)^2 + [\log(1+\omega_c^2\tau^2)]^2/4}, \text{(SE12)}$$

$$\rho_{xy}(B) = \rho_0 \omega_c \tau \frac{\frac{[\log(1+\omega_c^2\tau^2)]}{2}}{(\tan^{-1}\omega_c\tau)^2 + \frac{[\log(1+\omega_c^2\tau^2)]^2}{4}}, \text{(SE13)}$$

where $\rho_0 = [(S_{tot}/V)\sigma_0^{2D}]^{-1}$. In **Fig. S9**, the Hall resistivity $\rho_{xy}$ and the second derivative of $\rho_{xy}$ against $\omega_c\tau = \mu B$ were plotted. The plots showed good agreements with the qualitative features in the experimental results, including the characteristic kink structure in $\rho_{xy}$ (i.e., the peak in the second derivative) appearing at $\mu B \sim 0.7$.

**4.2 Hall resistance in graphene sponges**

Hall resistance curves of graphene sponges with various curvature radii at 2 K and 100 K were displayed in **Fig. S10**. The nonlinear Hall resistance curves were observed for all samples, and these curves resembled the theoretical curves, as shown in **Fig. S9**. This point was previously demonstrated by the $\mu B$ scaling of the Hall resistance curves in the wide range of carrier concentration in the graphene sponge EDLT device [S4, S12], indicating the inhomogeneous distributions of transverse magnetic fields on the 3D electron pathways. For 500 – 1000 nm sample, nonlinear behaviors in Hall resistivity curves were clearly detected at around $B = 1$ T. For 50 -150 nm sample, this behavior shifted to around $B = 2$ T and broadened against magnetic fields. Finally, for 25 – 50 nm sample, Hall resistivity curves became almost linear, and it gradually curved above 7 T. To elucidate the carrier mobility of graphene sponge as displayed in **Table 1**, we employed the conventional Hall mobility and mobility defined in the **Equation (SE13)**. The Hall mobility was calculated using electrical conductance normalized by the geometric dimension and Hall coefficient at the maximum magnetic field in each curve. The mobility in **Equation (SE13)** was derived from $\mu B \sim 0.7$ by a substitution of a value of the magnetic field where the second derivative of Hall resistance curves showed a peak.

### 4.3 Gate voltage dependence of the longitudinal resistance and conductance of graphene sponges

**Figure S5** showed the $V_G$ dependence of longitudinal resistance and conductance of the graphene sponges. The electrical conductance was calculated from the inverse of the longitudinal resistance. The electrical conductance showed ultrahigh values compared to the 2D CVD graphene[S6, S14] and decreased with the decrease of curvature radius. The conductance showed a minimum in the $V_G$ range from −0.5 to 0.5 V due to the ambipolar electronic states and no apparent enhancement of the conductance on/off ratio ($\sim 1.3 - 5$). Thus, the decrease of the electrical conductance with curvature radius was related to the enhancement of electron scattering by decreasing the curvature radius rather than the formation of the bandgap.

### 4.4 Capacitance of graphene sponges

**Figure S7** showed $V_G$ dependence of capacitance for the graphene sponges. The capacitance normalized by the projected area of graphene sponges showed high values (0.4 – 1 mF/cm$^2$ for the 500 – 1000 nm sample, 1 – 2.5 mF/cm$^2$ for the 50 – 150 nm sample, 2.3 – 3.6 mF/cm$^2$ for the 25 – 50 nm sample), which were 500~1000 times higher than that of 2D graphene EDLT (0 – 0.005 mF/cm$^2$) [S6]. Moreover, the quantum capacitance behavior [S6, S7, S15], associated with the linear density of states of Dirac fermion, was demonstrated in all samples.

### 4.5 Temperature dependence of electrical conductance of graphene sponges

Temperature dependence of electrical conductance of the graphene sponges showed $\log T$ curves at low temperatures due to both WL/WAL and EEI. With the increase in temperature, the $\log T$ curves of the conductance changed to the temperature-linear like curves for the 50 – 1000 nm samples, while the deviation from $\log T$ curves became smaller in the 25 – 50 nm sample, as shown in **Fig. 3**(a-c). In 2D graphene, the origin of positive temperature–conductance curve was clarified as the thermal excitation from the electron and hole puddle around the Dirac neutral point. The graphene sponge EDLT provided a rough estimation of the charge puddle within $\Delta V_G \sim 0.6\ V$ based on the $V_G$ – Hall resistance curves [S4]. From the surface area and the capacitance of 50 – 150 nm sample, the charge puddle was approximately estimated to be $n \sim 10^{11} - 10^{12}$ cm$^{-2}$, giving negligible temperature dependence up to room temperatures [S16]. Instead of the charge puddle, the reduction of the relaxation times with the increase in temperatures due to the effect of the electron-phonon scattering could be assumed from the suppression of the WL correction. In this case, the EEI correction in the diffusive limit ($\Delta G \propto \log T$) was gradually changed toward that in the ballistic one ($\Delta G \propto T$) with the increase in temperature [S17]. Alternatively, the local quasiperiodic potential, the circular pipe structure, and the high curvature induced pseudo magnetic fields may open the local energy gap. In this case, the electronic states consisting of the sum of massless and massive Dirac fermions could be sources of the $T$-linear resistivity.

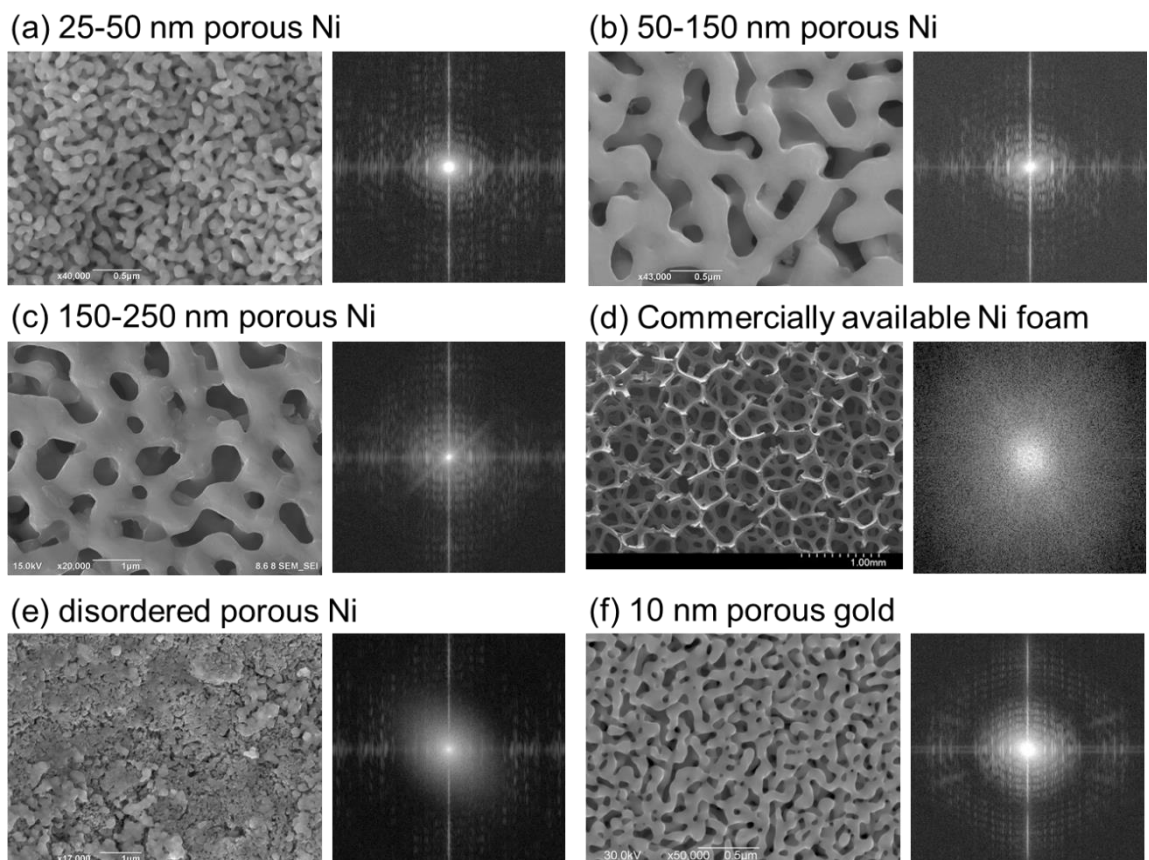

**Figure S1.** SEM images of (a-e) pore/ligament size, morphology and corresponding FFT patterns of porous Ni as CVD templates and (f) nanoporous gold with the corresponding FFT pattern. The porous Ni in (a-c) demonstrated similar FFT patterns with the nanoporous gold and dissimilar FFT patterns with the Ni foams in (d-e). The nanoporous gold was prepared by dealloying (Appl. Phys. Lett. 92, 251902 (2008).)

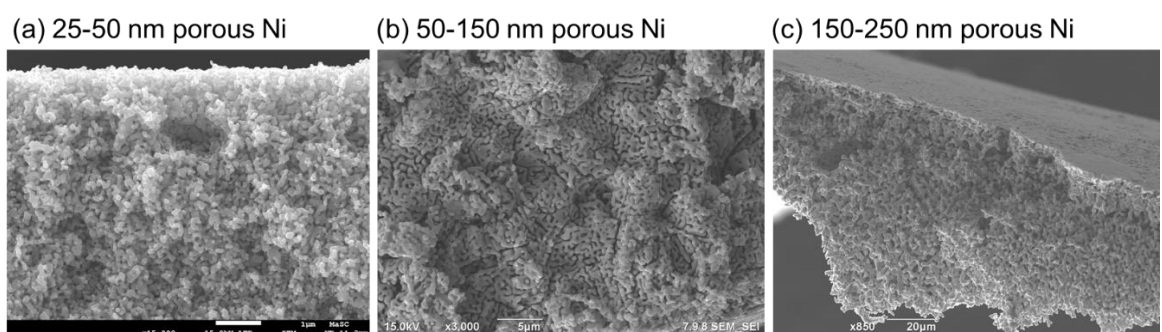

**Figure S2.** (a-c) Cross-sectional SEM images for the CVD templates. Bicontinuous structures were observed in the cross-section.

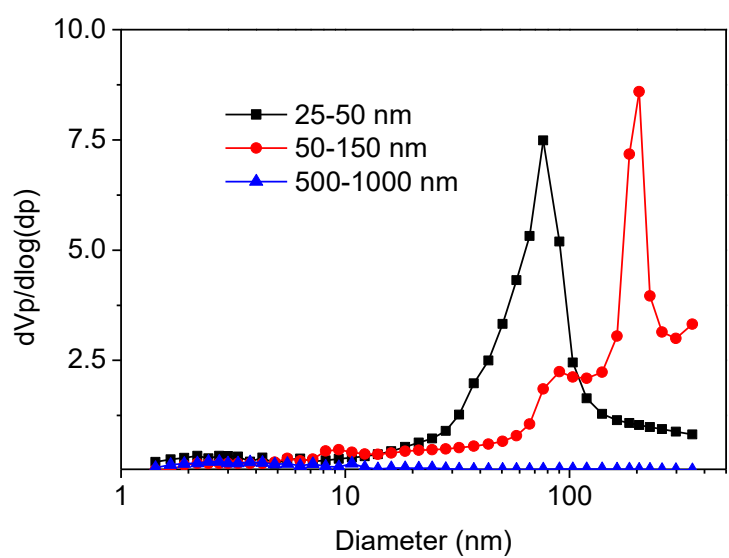

**Figure S3.** Barrett-Joyner-Hallender (BJH) pore distribution of graphene sponge. The peak pore diameter of 25–50 nm, 50–150 nm, and 500–1000 nm samples was 76.5 nm, 204.6 nm, and no peak pore diameter. The measurable limitation of this equipment was in the range of 0–350 nm.

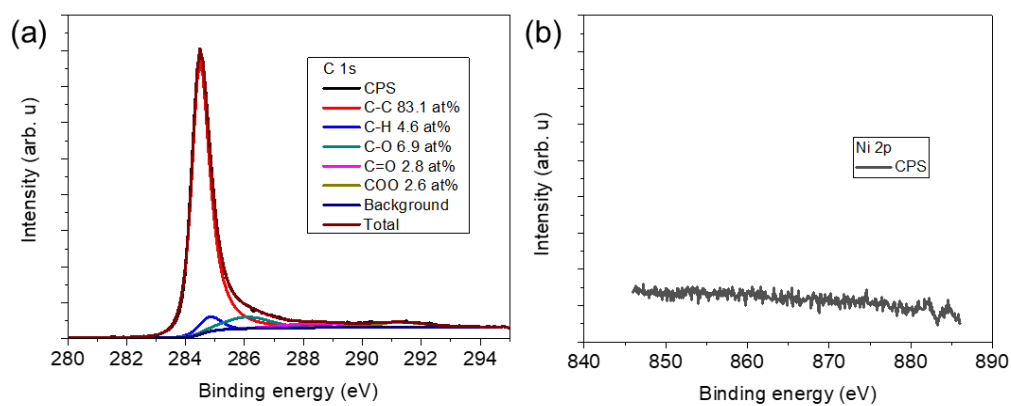

**Figure S4.** Typical XPS spectrum of graphene sponge. (a) C 1s and (b) Ni 2p spectra of the graphene with a 25–50 nm curvature radius. The residue Ni was less than 0.01 at.% for all graphene sponge samples.

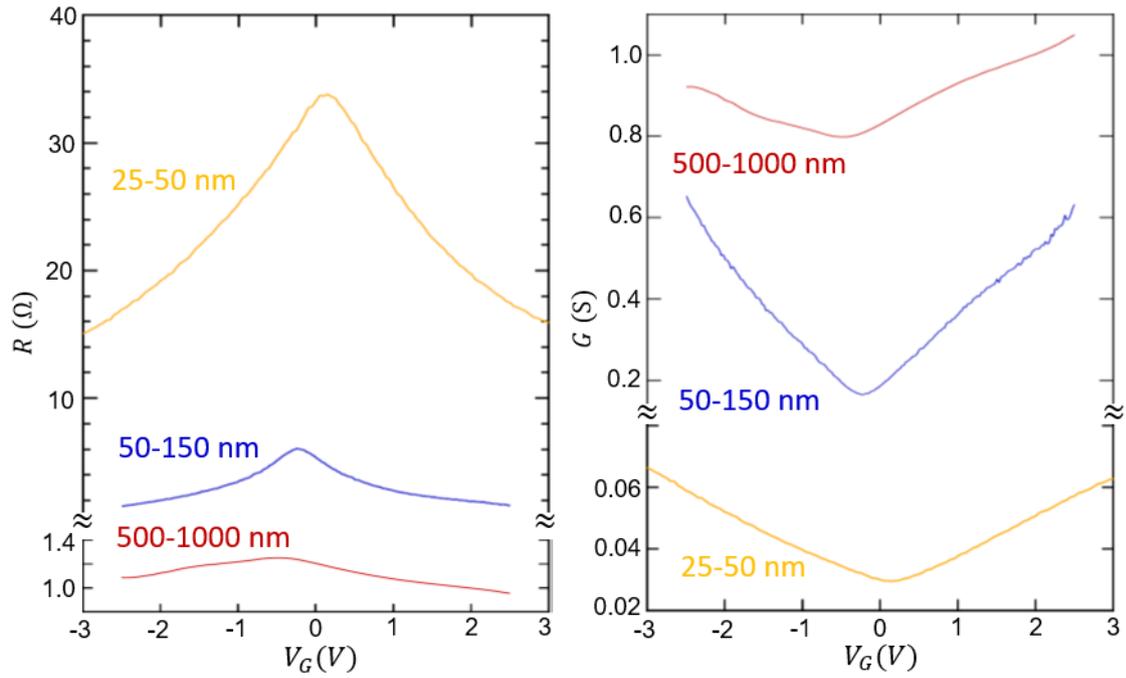

**Figure S5.** Longitudinal resistance and conductance against the gate voltage $V_G$ for the graphene sponge EDLT with 25 – 1000 nm curvature radii at room temperature.

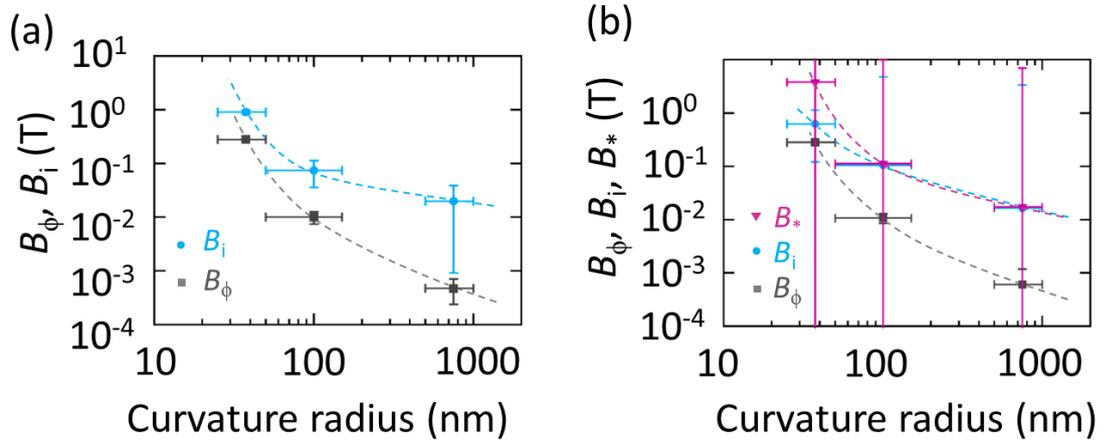

**Figure S6.** Analysis of weak localization in magnetoconductance curves.

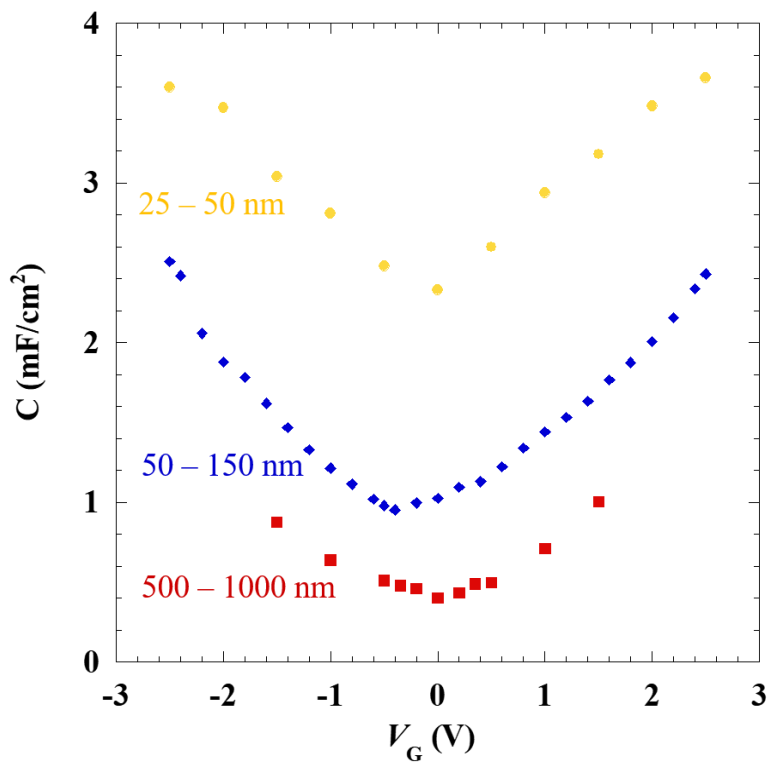

**Figure S7.** The capacitance of graphene sponge. Gate voltage dependence of capacitance for graphene sponge samples with various curvature radii.

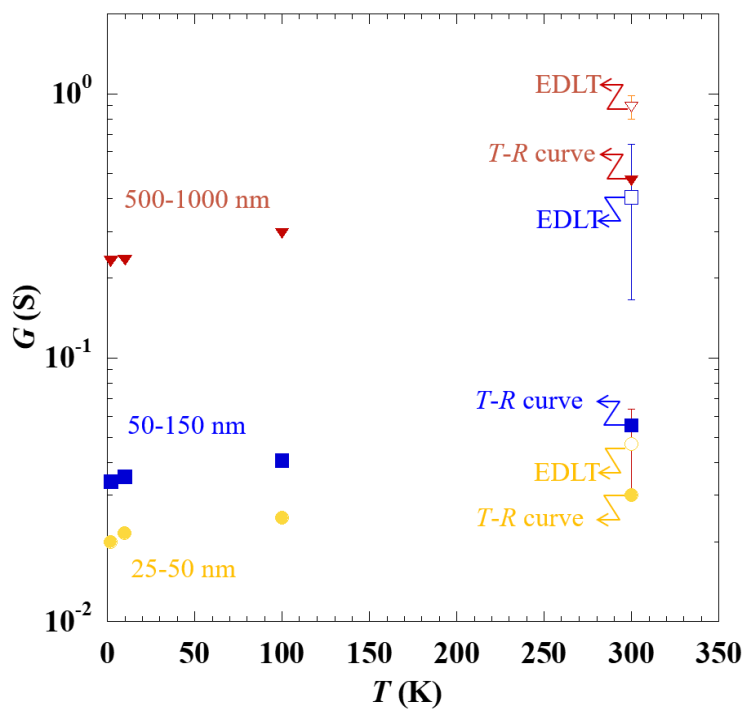

**Figure S8.** The electrical conductance of graphene sponges. Open symbols were the electrical conductance calculated from the inverse of the electrical resistance of the graphene sponge electric double layer transistor (EDLT). Closed symbols were those from the temperature

dependence of electrical resistance (*T-R* curves). The electrical conductance from 2 K to 300K was systematically suppressed with a decrease in curvature radius of graphene sponges.

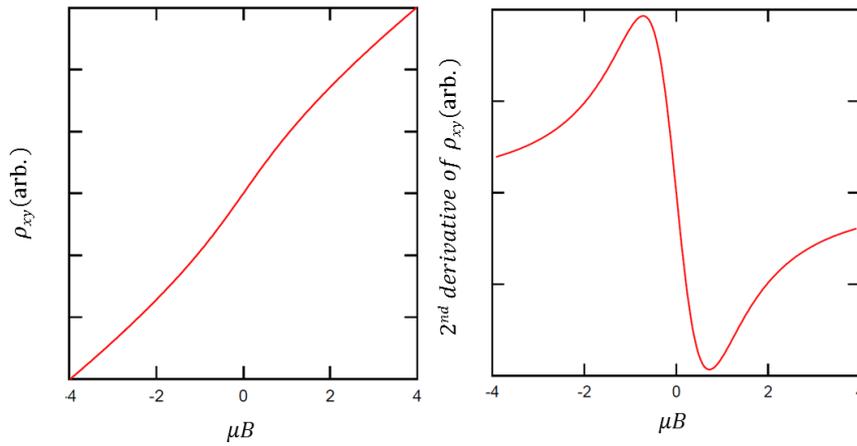

**Figure S9.** Theoretical magnetic field (*B*) dependence of the Hall resistivity ($\rho_{xy}$) and second derivative of $\rho_{xy}$ for the graphene sponge using **Equation (SE13)**. The peak positions at the X-axis were ±0.7.

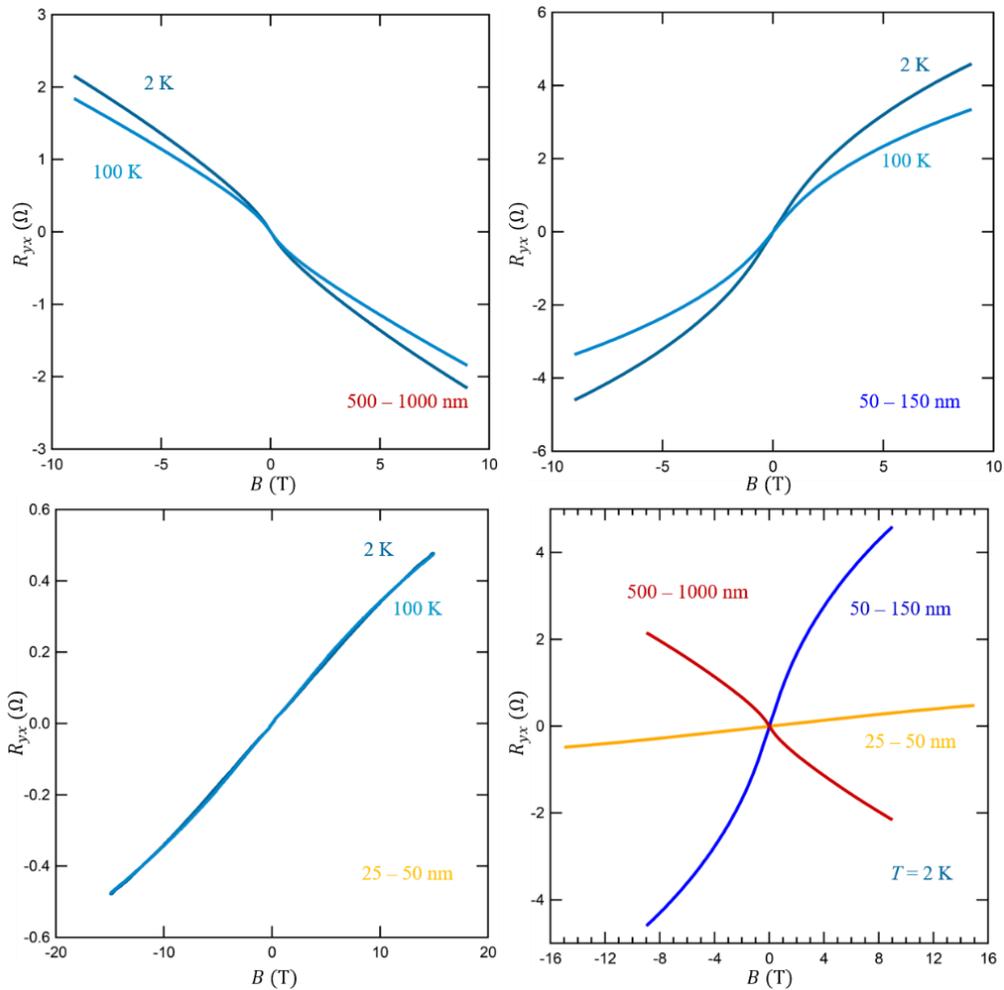

**Figure S10.** Hall resistivity curves of graphene sponge. Hall resistivity curves for various curvature radius at 2 K and 100 K were displayed. While non-linear Hall resistivity curves were confirmed for all samples, the curves were gradually changed to linear lines with the decrease in the curvature radius.

**Table S1.** Raman spectra measurements of the graphene sponge. The unit of the spectra and line width is cm$^{-1}$.

|  | D band Line width | G band Line width | D' band Line width | 2D band Line width | $I_D/I_G$ | $I_{2D}/I_G$ |
|---|---|---|---|---|---|---|
| 25–50 nm graphene sponge | 1351 21 | 1580 14 | 1618 16 | 2700 31 | 0.52 | 4.73 |
| 50–150 nm graphene sponge | 1350 21 | 1585 14 | - - | 2707 22 | 0.20 | 4.76 |
| 500–1000 nm graphene sponge | 1357 20 | 1588 10 | - - | 2713 18 | 0.04 | 4.64 |